\documentclass[%
 reprint,
superscriptaddress,
nobibnotes,
 amsmath,amssymb,
 aps,
prl,
citeautoscript,
floatfix,
showkeys
]{revtex4-2}

\usepackage[T1]{fontenc} 
\usepackage[]{graphicx}
\usepackage[utf8]{inputenc}
\usepackage{blindtext}
\usepackage{lmodern}%
\usepackage{hyperref}%
\usepackage{xcolor}%
\usepackage{placeins}
\usepackage{amsmath,amssymb}
\usepackage{bm}
\usepackage{tabularx}
\usepackage{booktabs}
\usepackage{notes2bib}
\usepackage[normalem]{ulem}
\usepackage[normalem]{ulem}
\usepackage{placeins}
\usepackage{amssymb}

\newcommand{\beq}{\begin{equation}\begin{aligned}}
\newcommand{\eeq}{\end{aligned}\end{equation}}

\newcommand{\ECA}{EuCd\ensuremath{_2}As\ensuremath{_2}}

\newcommand{\CPS}{CrPS$_4$}

\newcommand{\Tn}{\ensuremath{T_{\rm N}}}

\newcommand{\neel}{\emph{Néel}}
\newcommand{\vdW}{\emph{van der Waals}}

\newcommand{\SiOTwo}{SiO\textsubscript{2}}

\newcommand{\VG}{\ensuremath{V_{\rm {G}}}}
\newcommand{\Vth}{\ensuremath{V_{\rm TH}}}
\newcommand{\dd}{{\rm d}}

\newcommand{\figref}[2]{\ref{#1}\textsf{#2}}

\newcommand{\ie}{\emph{i.e.},}
\newcommand{\eg}{\emph{e.g.},}

\newcommand{\dqmp}{Department of Quantum Matter Physics, University of Geneva, 24 Quai Ernest Ansermet, CH-1211 Geneva, Switzerland}
\newcommand{\gap}{Department of Applied Physics, University of Geneva, 24 Quai Ernest Ansermet, CH-1211 Geneva, Switzerland}
\newcommand{\KW}{Research Center for Functional Materials, NIMS, 1-1 Namiki, Tsukuba 305-0044, Japan}
\newcommand{\TT}{International Center for Materials Nanoarchitectonics, NIMS, 1-1 Namiki, Tsukuba 305-0044, Japan}
\newcommand{\modena}{Dipartimento di Scienze Fisiche, Informatiche e Matematiche, University of Modena and Reggio Emilia, IT-41125 Modena, Italy}
\newcommand{\cnr}{Centro S3, CNR Istituto Nanoscienze, IT-41125 Modena, Italy}

\definecolor{linkcol}{rgb}{0,0,0.4}
\definecolor{citecol}{rgb}{0.5,0,0}
\definecolor{harvardcrimson}{rgb}{0.79, 0.0, 0.09}
\definecolor{lava}{rgb}{0.81, 0.06, 0.13}

\hypersetup{
	bookmarksopen=true,
	pdftitle="Magnetism-induced band-edge shift as mechanism for magnetoconductance in CrPS4 transistors",
	pdfauthor="F.Wu et al",
	pdfsubject="", 
	pdfstartview={FitH}, 
	pdfmenubar=true, 
	pdfhighlight=/O, 
	colorlinks=true, 
	pdfpagemode=UseNone, 
	pdfpagelayout=SinglePage, 
	pdffitwindow=true, 
	linkcolor=harvardcrimson, 
	citecolor=harvardcrimson, 
	urlcolor=lava
}

	\begin{abstract} 
	Transistors realized on 2D antiferromagnetic semiconductor \CPS\ exhibit large magnetoconductance, due to magnetic-field-induced changes in magnetic state. The microscopic mechanism coupling conductance and magnetic state is not understood. We identify it by analyzing the evolution of the parameters determining the transistor behavior --carrier mobility and threshold voltage-- with temperature and magnetic field. For temperatures $T$ near the \neel\ temperature \Tn, the magnetoconductance originates from a mobility increase due to the applied magnetic field that reduces spin fluctuation induced disorder. For $T<<\Tn$, instead, what changes is the threshold voltage, so that increasing the field at fixed gate voltage increases the density of accumulated electrons. The phenomenon is explained by a conduction band-edge shift correctly predicted by \emph{ab-initio} calculations. Our results demonstrate that the bandstructure of \CPS\ depends on its magnetic state and reveal a mechanism for magnetoconductance that had not been identified earlier.
	\end{abstract}

\begin{document}
	
	
	\title{Magnetism-induced Band-edge Shift as Mechanism for\texorpdfstring{\\}{} Magnetoconductance in \texorpdfstring{\CPS}{CrPS4} Transistors}

	\author{Fan Wu} 
 \email{fan.wu@unige.ch}
 \affiliation{\dqmp}
 \affiliation{\gap}
	\author{Marco Gibertini} 
 \affiliation{\modena}
 \affiliation{\cnr}
 \author{Kenji~Watanabe} 
	\affiliation{\KW}
	\author{Takashi~Taniguchi} 
	\affiliation{\TT}
 \author{Ignacio Gutiérrez-Lezama} 
	\author{Nicolas~Ubrig}
	\email{nicolas.ubrig@unige.ch}
 \author{Alberto F. Morpurgo}
 \email{alberto.morpurgo@unige.ch}
 \affiliation{\dqmp}
 \affiliation{\gap}

 	\date{\today}

	\maketitle

\maketitle

Many fascinating phenomena observed in atomically thin 2D magnets arise from the interplay between the magnetic state of the material and processes characteristic of semiconductor physics~\cite{burch_magnetism_2018,gibertini_magnetic_2019,mak_probing_2019}. Examples include giant magnetoconductance in tunnel barriers~\cite{klein_probing_2018,song_giant_2018,wang_very_2018,kim_one_2018}, the electrostatic control of magnetic phase boundaries~\cite{huang_electrical_2018,jiang_controlling_2018,wang_electric-field_2018,verzhbitskiy_controlling_2020}, or the dependence of the wavelength of emitted light on the magnetic state~\cite{wilson_interlayer_2021}. More predictions have been made and remain to be validated, such as the realization of gate-tunable half-metals~\cite{li_half-metallicity_2014,gong_electrically_2018,deng_two-dimensional_2021}, or the possibility of engineering skyrmion-like textures to control electron dynamics in magnetic moiré stacks ~\cite{tong_skyrmions_2018,akram_moire_2021,mak_semiconductor_2022}. Assessing the validity of these predictions and understanding in detail the phenomena already observed is however difficult, because most 2D magnetic semiconductors studied so far have extremely narrow electronic bandwidth~\cite{wang_electronic_2011,zheng_ab_2019}. These materials are therefore expected to behave differently from conventional semiconductors, because in narrow bandwidth systems electron-electron interactions and disorder typically play a dominant role. As a result, it is not clear a priori whether conventional band theory can be used to describe the properties of 2D magnetic semiconductors as it does for common semiconducting compounds.

\begin{figure}
	\centering
	\includegraphics[width=\linewidth]{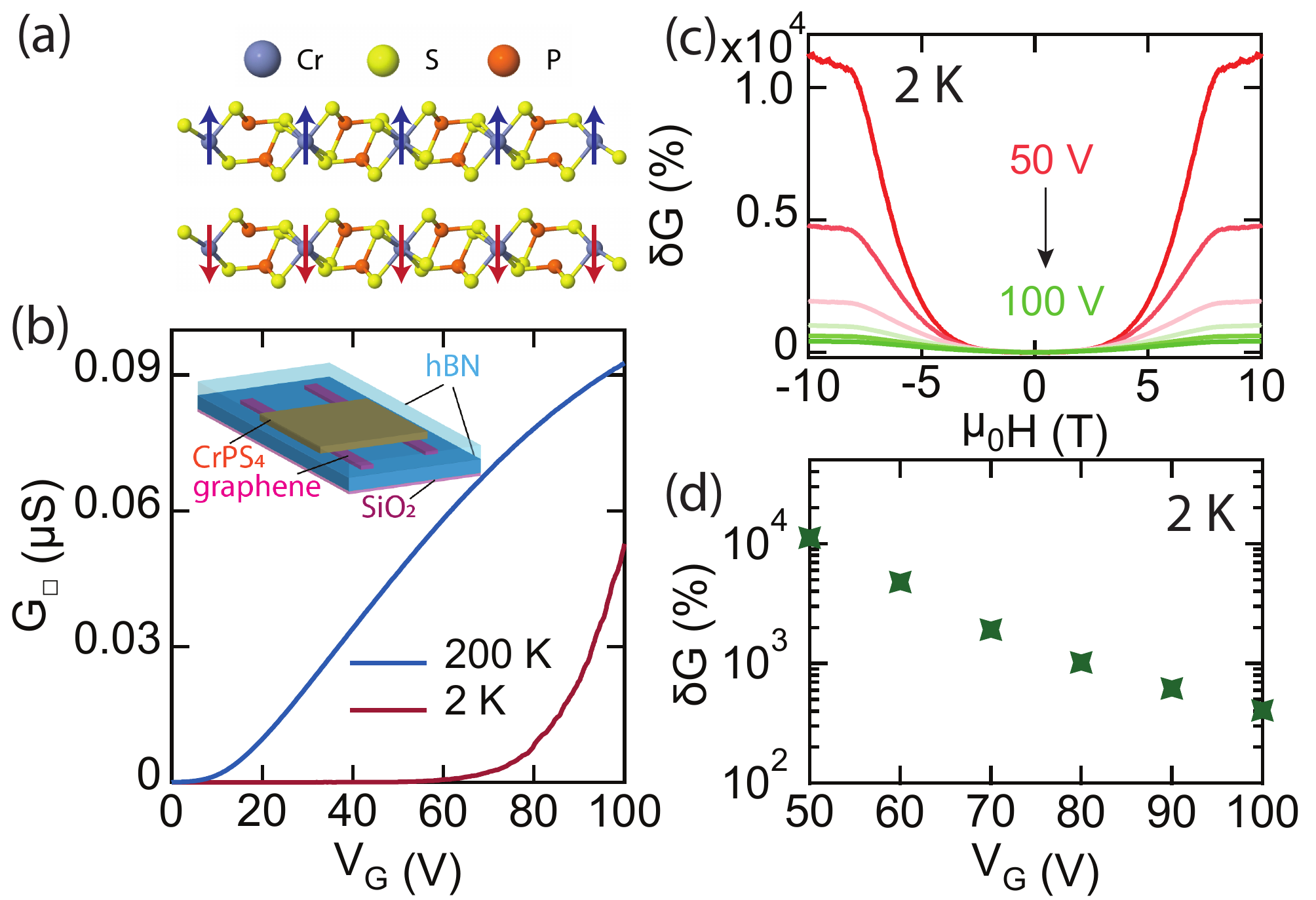}
	\caption{ (\textsf{a}) Crystal structure of layered \CPS; the blue, yellow, and orange spheres represent Cr, S, and P atoms, respectively. \CPS\ has an antiferromagnetic ground state (A-type) formed by individual layers uniformly magnetized in the out-of-plane direction (\Tn~$\approx$~35~K)~\cite{peng_magnetic_2020,calder_magnetic_2020,son_air-stable_2021}. Panels (\textsf{b})-(\textsf{d}) summarize the key features of \CPS-transistors reported in Ref. ~\cite{wu_gate-controlled_2023}. (\textsf{b}) Transfer curves ($G_{\square}$-vs-\VG) of a 10~nm thick \CPS\ FET device at 200 and 2~K. The inset shows the device schematics. A hBN-encapsulated \CPS\ multilayer --with thickness ranging from 6 to 10 nm in different devices-- is contacted by graphene stripes and the gate voltage \VG\ is applied across a 285~nm \SiOTwo\ insulating layer (for experimental details about the precise configuration of the device and its fabrication see Ref. ~\cite{wu_gate-controlled_2023}). (\textsf{c}) Magnetoconductance ($\delta G = \frac{G(\mu_0H)-G(0~T)}{G(0~T)}$) measured at $T$~=~2~K for fixed \VG\ values in the 50-100~V range, in 10~V steps. (\textsf{d}) The magnetoconductance at $\mu_0H$~=~10~T and $T$~=~2~K depends exponentially on \VG.}
	\label{fig:1}
	
\end{figure}

The effect of the narrow bandwidth also has a pronounced impact on the transport properties, which is why the realization of field-effect transistors enabling the controlled and systematic investigation of transport as a function of electron density has proven difficult on 2D magnetic semiconductors~\cite{klein_probing_2018,song_giant_2018,wang_very_2018,ghazaryan_magnon-assisted_2018,kim_one_2018,soler-delgado_probing_2022}. Only recently, first materials have been identified with a bandwidth of 1~eV or larger~\cite{zhuang_density_2016,telford_layered_2020,liu_vapor_2020,susilo_band_2020}, allowing transistors to operate well below the magnetic critical temperature and enabling systematic studies of magnetotransport ~\cite{wu_quasi-1d_2022,lebedev_electrical_2023,wu_gate-controlled_2023}. In devices based on multilayers of the most recently reported compound --the \vdW\ antiferromagnetic semiconductor \CPS\ (see Fig.~\figref{fig:1}{a,b})-- the magnetoconductance has been found to be very large (reaching up to $\approx~10000$~\%) and strongly dependent on the gate voltage (as we reproduce in Fig.~\figref{fig:1}{c,d} for convenience)~\cite{wu_gate-controlled_2023}. It has been clearly established that the measured magnetoconductance is determined by how the magnetic state of \CPS\ evolves when a magnetic field is applied, but its precise microscopic origin could not be identified. 

Here, we analyze the transport properties of transistors realized on the layered antiferromagnetic semiconductor \CPS, and demonstrate that the large, gate-tunable magnetoconductance observed at low temperature originates from the dependence of the bandstructure on the magnetic state. More specifically, we find that the dominant mechanism responsible for the observed magnetoconductance at temperature $T\simeq T_N$ and at $T\ll T_N$ is different. In the vicinity of \Tn, transport is dominated by the effect of spin-fluctuations that limit the mobility of charge carriers, as it commonly happens in magnetic conductors. For $T\ll\Tn$, instead, the mobility does not depend on temperature or magnetic field. In this regime, the large observed magnetoconductance originates entirely from the shift of the conduction band-edge to lower energy, which results --at fixed applied gate voltage-- in an increase in the density of accumulated electrons. We show that such an effect is expected from \emph{ab-initio} calculations, which predict the magnitude of the band-edge shift between the antiferromagnetic state at $H=0$ and the ferromagnetic state at high field to be comparable to the value that we estimate from experimental data. These conclusions demonstrate that the electronic bandstructure of \CPS\ does depend on the magnetic state, and that this dependence results in a microscopic mechanism that can generate very large, gate-tunable magnetoconductance in magnetic semiconductors when the Fermi level is sufficiently close to the conduction band-edge.

The overall experimental phenomenology of the magnetotransport response of \CPS\ transistors has been very recently reported in Ref.~\cite{wu_gate-controlled_2023}. That work presents all key experimental observations, but does not discuss the physical mechanisms responsible for the observed magnetotransport, which remains to be determined. Here we focus on the analysis of the transistor response to identify these mechanisms. We gain new insight by analyzing systematically the evolution with temperature and applied magnetic field of the parameters that determine the transistor operation. Specifically, the square conductance of a field-effect transistor 
\begin{eqnarray}
	G_{\square} = \mu \cdot C_{\rm } \cdot (\VG - V_{\rm TH}). 
	\label{eq:FET}
\end{eqnarray}
is a function of the gate voltage, \VG, and depends on three parameters, each carrying information about different microscopic properties~\cite{sze_physics_2006}. The capacitance to the gate electrode $C$ is determined by the in-series connection of the geometrical capacitance $C_{\rm G}$ and of the quantum capacitance $C_{\rm Q}$ ($1/C=1/C_{\rm G}+1/C_{\rm Q}$), with $C_{\rm Q}$ proportional to the density of states. Measuring $C$ may therefore allow the density of states in the different magnetic phases of the material to be determined. In the devices considered here, however, the geometrical capacitance is small and the effect of the quantum capacitance is negligible (\ie\ $C=C_{\rm G}$). The threshold voltage \Vth\ is determined by the energetic position of the conduction band-edge ($V_G=V_{\rm TH}$ corresponds to having the Fermi level $E_F$ aligned with the conduction band-edge) and variations in \Vth\ with magnetic field can therefore signal a change in bandstructure. Finally, the carrier mobility $\mu$ provides information about scattering processes and disorder mechanisms affecting charge carriers. As well-established, for electrons in magnetic conductors, spin disorder is commonly found to be a dominant mechanism determining the mobility, with better spin alignment leading to higher mobility values. 

\begin{figure}
	\centering
	\includegraphics[width=\linewidth]{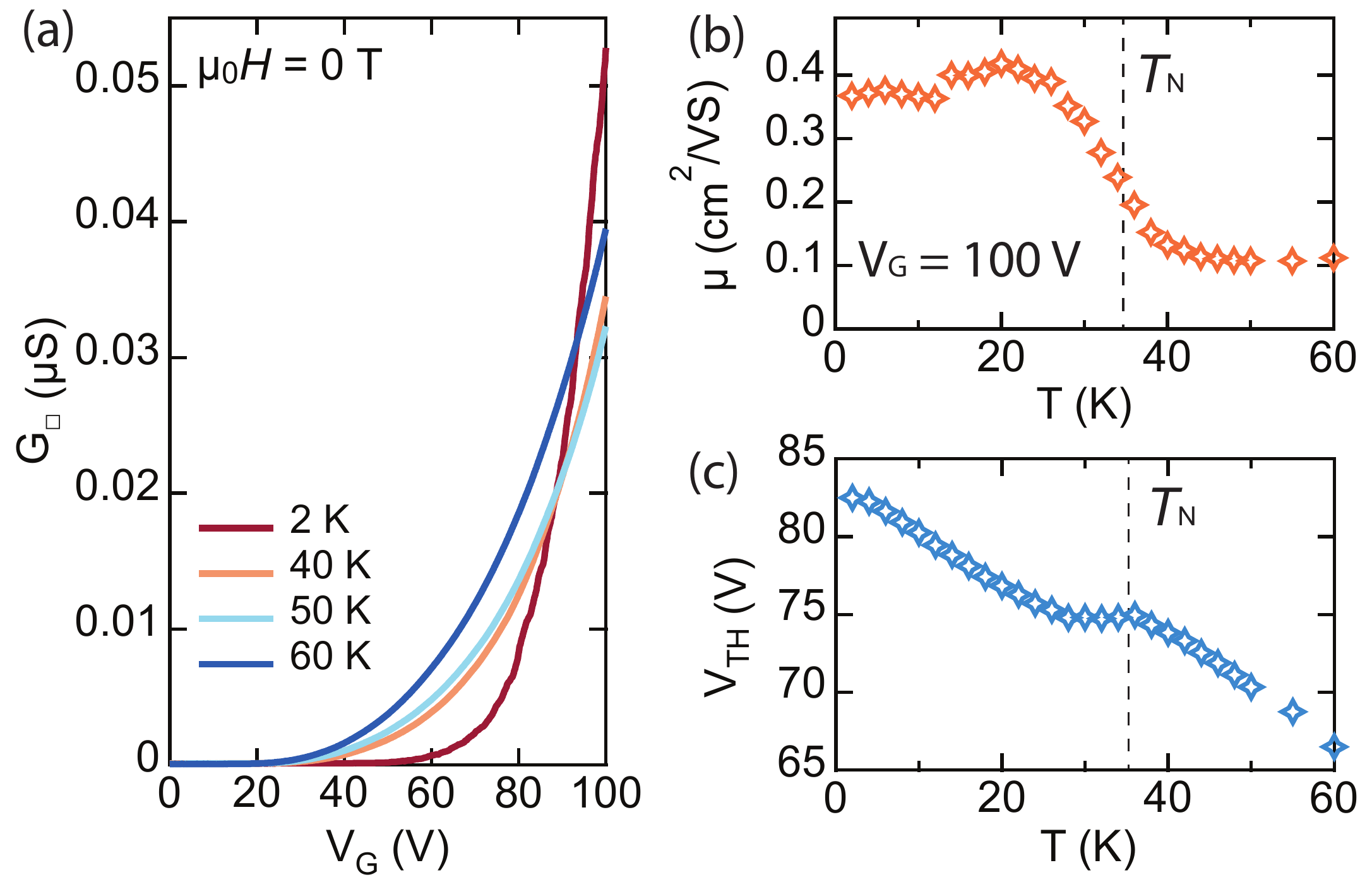}
	\caption{ (\textsf{a}) Transfer curves of a transistor based on a 10~nm \CPS\ multilayer, measured at $\mu_0H =0$, at $T=$~2, 40, 50, and 60~K (as indicated in legend). (\textsf{b}) $T$-dependence of the field-effect mobility $\mu$ extracted from the transconductance (see Equation~\ref{eq:FET}) at \VG\ of +100~V, exhibiting a fourfold increase as $T$ is lowered below \Tn. (\textsf{c}) $T$-dependence of the the threshold voltage, obtained by extrapolating the square conductance in \textsf{a} to zero. All data presented in the main text have been measured on this device. Additional data from other devices are shown in the supporting information.}
	\label{fig:2}
	
\end{figure}

To analyze the properties of \CPS\ transistors, we start by looking at the transfer curves ($G_{\square}$-vs-$V_G$; details about the device fabrication can be found in Ref~\cite{wu_gate-controlled_2023}). Fig.~\figref{fig:2}a shows the transfer curves measured for different values of $T$ ranging from above $\Tn~\simeq~35$~K to 2~K for a 10~nm thick device (all data shown in the main text has been measured on this same device; data from additional devices can be found in the supporting information). Fig.~\figref{fig:2}b shows that the mobility extracted from the transconductance ($\mu = \frac{1}{C_{\rm }} \frac{\partial G_{\square}}{\partial \VG}$) increases by a factor of 4 as $T$ is decreased below \Tn, and eventually saturates at low $T$. Concomitantly, the threshold voltage (obtained by extrapolating to zero the square conductance measured as a function of \VG) increases gradually by approximately 10~V as $T$ is lowered from 60~K to 2~K (Fig.~\figref{fig:2}c). Both trends are easily understood in terms of established concepts. The mobility starts increasing at \Tn, because ordering of the spins in the magnetic states suppresses spin fluctuations, thereby effectively decreasing the magnitude of disorder experienced by electrons. The modest increase in \Vth\ upon cooling can be attributed to the freeze out of charge carriers into the dopants where they originate from, and can be expected for \CPS\ that is indeed unintentionally doped. 
\begin{figure}
	\centering
	\includegraphics[width=\linewidth]{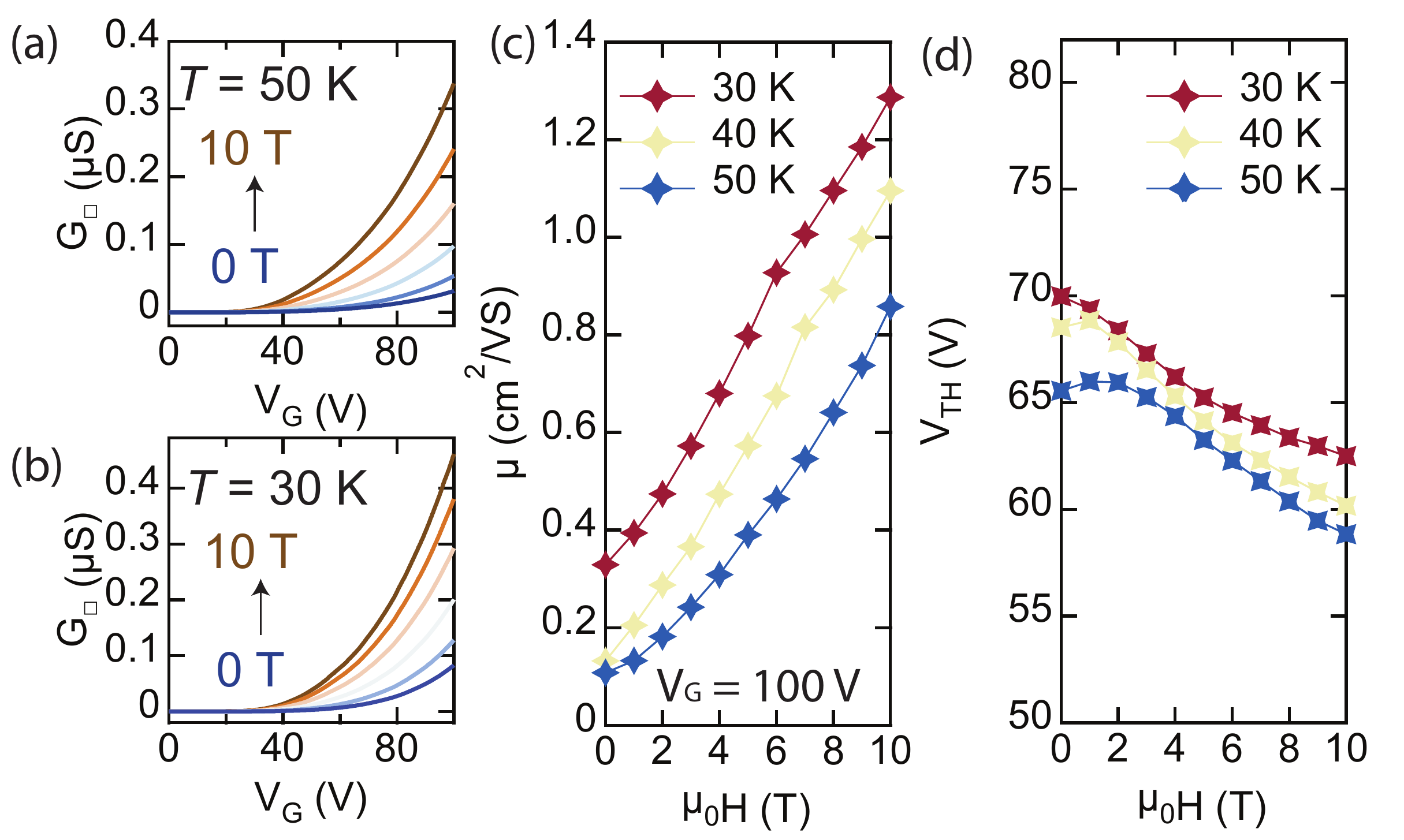}
	\caption{(\textsf{a, b}) Transfer curves of the device discussed in Fig.~\ref{fig:2}, for different values of $\mu_0H$ between 0 and 10~T, at $T$~=~50 and $T$~=~30~K, respectively above and below \Tn. (\textsf{c}) Magnetic field dependence of $\mu$ obtained from the device transconductance at \VG~=~+100~V, for three different temperatures close to \Tn (30~K, 40~K, and 50~K for the red, yellow and blue symbols), showing an increase of nearly one order of magnitude. (\textsf{d}) Dependence of \Vth\ on magnetic field at the measured at the same three temperatures, showing a variation of less than 10~\% as $\mu_0H$ is increased from 0 to 10~T. }
	\label{fig:3}
\end{figure}

These conclusions are fully consistent with the dependence of $\mu$ and \Vth\ on applied magnetic field $\mu_0H$, for temperatures $T$ near (above or just below) \Tn. Transfer curves at two different temperatures (50~K and 30~K) are shown in Fig.~\figref{fig:3}{a} and \figref{fig:3}{b} for different values of applied magnetic field, ranging from 0 to 10~T. For both temperatures, the magnetic field causes the conductance to increase by 5-to-10 times (see, \eg\ $G_{\square}$ at \VG~=~+100~V, in Fig.~\figref{fig:3}{a} and \figref{fig:3}{b}), with the increase originating from the transconductance $\frac{\partial G_{\square}}{\partial \VG}$. Indeed, Fig.~\figref{fig:3}{c} shows that $\mu$ increases by nearly the same amount as the conductance as $\mu_0 H$ is increased from 0 to 10~T. The mobility is calculated at the largest value of \VG\ reached in the experiments, \VG~=~+100~V, to ensure that the transistors operate in the linear regime. The threshold voltage changes by less than 10~\% (Fig.~\figref{fig:3}d). This is again expected, because the application of the magnetic field aligns the spins and reduces the disorder experienced by charge carriers (associated to spin fluctuations), for both $T>\Tn$ in the paramagnetic state of \CPS, and just below \Tn\ (see below for a discussion of the change in \Vth). Finding that for $T$ close to \Tn\ magnetotransport in \CPS\ transistors can be understood in terms of established paradigms is important, because it gives us confidence that our approach to identify the microscopic origin of the magnetoconductance is correct.

\begin{figure}
	\centering
	\includegraphics[width=\linewidth]{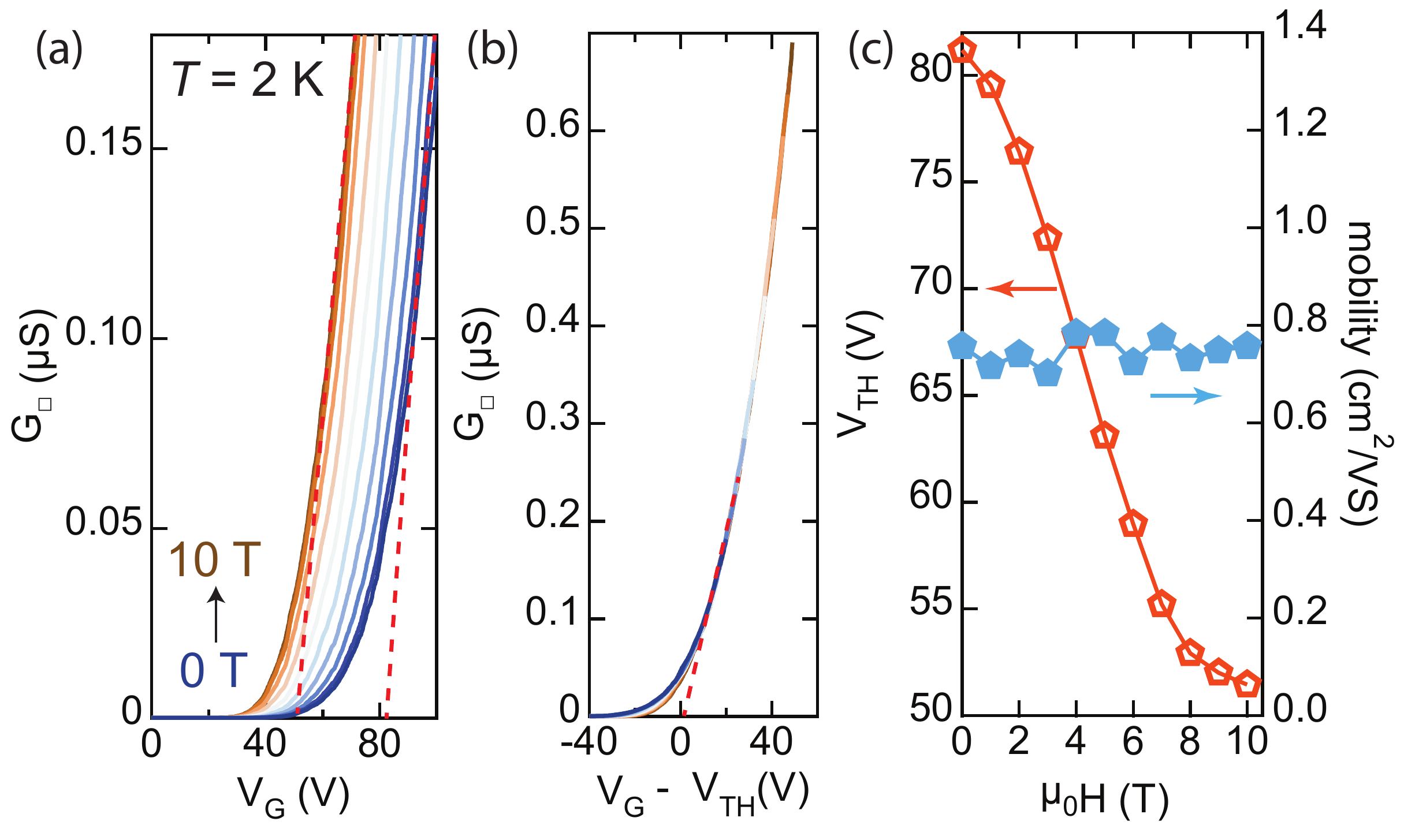}
	\caption{(\textsf{a}) Transfer curves of the same device whose data are presented in Fig.~\ref{fig:2} and \ref{fig:3}, measured at $T$~=~2~K (\ie\ $T<< \Tn$) for different, fixed values of $\mu_0H$ ranging from 0 to 10~T. The red dashed lines show the extrapolation to zero of the square conductance, from which we determine \Vth. (\textsf{b}) The same transfer curves as in (\textsf{a}) plotted versus $\VG-V_{TH}(\mu_0H)$ collapse on top of each other. (\textsf{c}) Magnetic field dependence of \Vth\ (open orange symbols) and $\mu$ (light blue symbols) extracted from panel (\textsf{a}): the applied magnetic field causes a substantial shift of \Vth\ (nearly 30~V), leaving the mobility unchanged (identical behavior is seen in multiple devices; see supporting information). }
	\label{fig:4}
\end{figure}

Having established that magnetotransport near \Tn\ is determined by the influence of spin fluctuations on electron mobility, we now discuss the magnetic field dependence of the conductance for $T\ll\Tn$, whose behavior is strikingly different. To illustrate the difference, Fig.~\figref{fig:4}{a} shows the transfer curves of a \CPS\ transistor measured at $T=2$~K for different applied magnetic fields, from which it is apparent that changing the magnetic field induces a large shift in \Vth. When each transfer curve is shifted by the corresponding threshold voltage --\ie\ when plotting the data as a function of $\VG-V_{\rm TH}(H)$-- all curves collapse. The collapse directly implies that the transconductance (measured at a same value of $\VG-V_{\rm TH}(H)$) is independent of the applied magnetic field, so that the mobility is also magnetic field independent. It follows that over the entire range of magnetic fields explored, the dependence of the conductance on field originates from the shift of \Vth\ with $H$, as illustrated quantitatively by the plots of \Vth\ and $\mu_0H$ in Fig.~\figref{fig:4}{c}. The threshold voltage downshifts by 30 V as the magnetic field is increased from 0 to 8~T (\ie\ the spin-flip field of \CPS\ at $T=2$~K) and saturates past that, while the mobility remains constant at $\mu\simeq 0.8$~cm$^2$/Vs. Even though different devices show somewhat different mobility values (reaching up to $\mu\simeq 6 $~cm$^2$/Vs; see supporting information), the observed downshift in threshold voltage is robust and virtually identical in all cases (if normalized to the value of the gate capacitance).

\begin{figure}
	\centering
	\includegraphics[width=\linewidth]{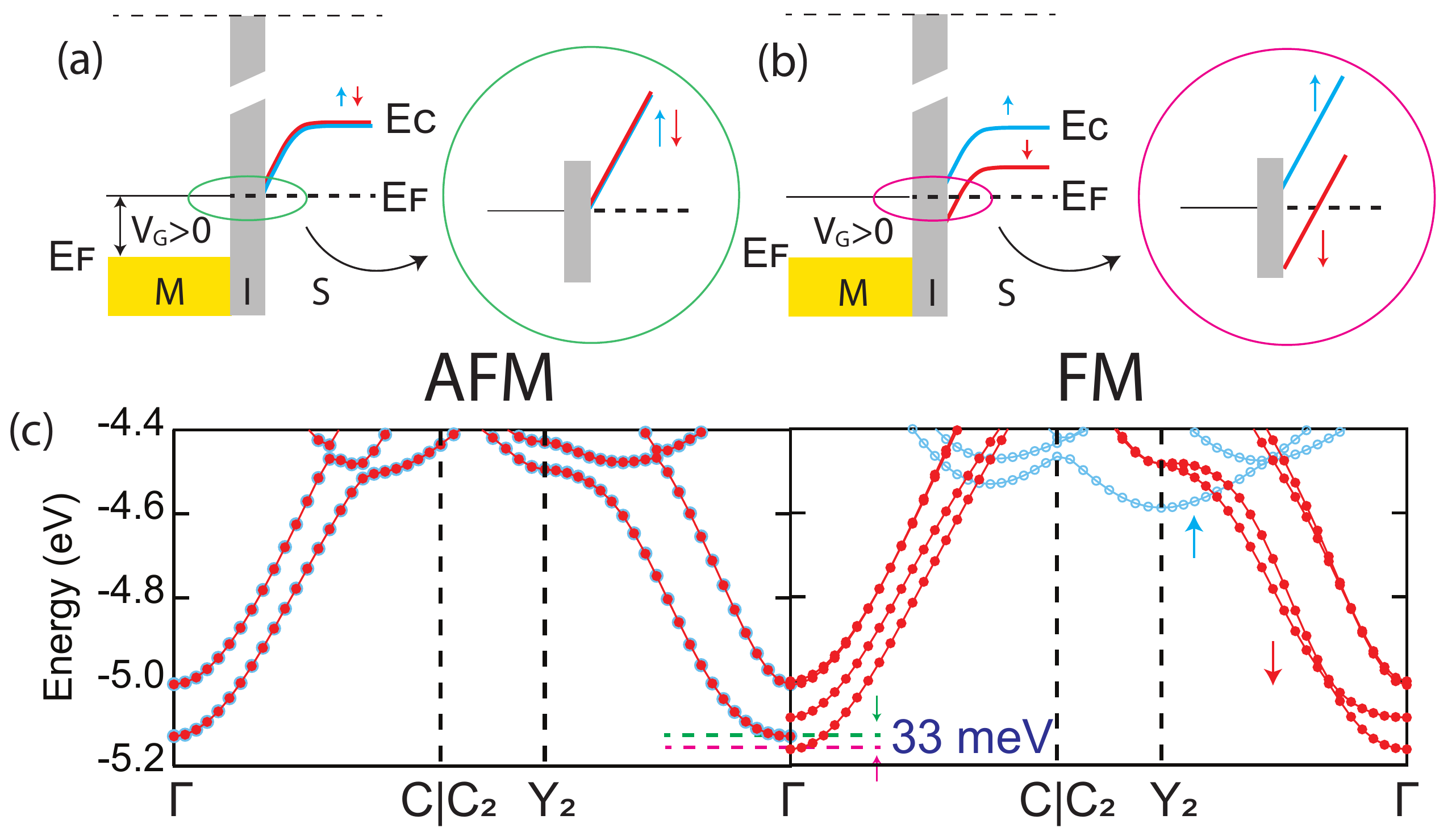}
	\caption{(\textsf{a, b}) Schematic representation of the energy band diagram for our \CPS\ transistors, illustrating the shift of the Fermi level \textsf{E}$_{\mathsf F}$ relative to the conduction band-edge \textsf{E}$_{\mathsf C}$ as the magnetic state changes from antiferromagnetic (AFM, left) to ferromagnetic (FM, right) ground state (the blue and red lines represent the spin-up and spin-down conduction bands). (\textsf{c}) First-principles calculation of the bandstructure of \CPS\ bilayers in the AFM and FM states. Blue and red colors represent the spin-up and spin-down states, respectively. The conduction band-edge (green and purple dashed lines) in the FM phase is 33~meV lower than in the AFM phase.}
	\label{fig:5}
\end{figure}

The very different nature of magnetotransport for $T\approx \Tn$ and for $T\ll\Tn$ originates from the distinct microscopic mechanisms that cause the conductance to depend on applied field in the two temperature regimes. At high temperature magnetoconductance occurs because the magnetic field decreases spin-induced disorder experienced by charge carriers. At low temperature, however, the same mechanism becomes inactive, because the antiferromagnetic state of \CPS\ is fully developed, and spins are already ordered at $H=0$. Finding that the magnetoconductance originates from the shift in threshold voltage indicates that what changes upon applying the magnetic field is the energetic position of the conduction band-edge (see schematic illustration in Fig.~\figref{fig:5}{a} and \figref{fig:5}{b}). Therefore, increasing $H$ at fixed \VG\ leads to an increase in the density $\Delta n$ of accumulated electrons ($\Delta n= C_{\rm } \cdot (\VG - \Delta V_{\rm TH})/e$) contributing to transport, and results in a larger conductance. The effect is sizable, because a shift in \Vth\ of 30~V corresponds to a variation in electron density in the transistor channel of $\Delta n~=~2.2 \cdot 10^{12}$~cm$^{-2}$. This mechanism also explains the exponential dependence of the magnetoconductance on gate voltage observed experimentally (see Fig.~\ref{fig:1}d). Such a dependence is observed when transistors operate in the sub-threshold regime, with \VG\ large enough to accumulate mobile carriers, but still such that $\VG < \Vth$. A shift in \Vth\ induced by the magnetic field is then effectively equivalent to changing \VG\ in a regime where the conductance depends exponentially on gate voltage (the exponential dependence of the current on gate voltage is why one commonly defines the subthreshold swing $S=\dd \VG/\dd (\log{I})$ to characterize the sub-threshold device behavior).


To further substantiate the validity of our conclusion, we performed \emph{ab-initio} calculations of the bandstructure of \CPS\ with respectively antiferromagnetic and ferromagnetic spin configurations (corresponding to the experimental situations at $\mu_0H=0$ and $\mu_0H > 8$~T), to determine the position of the conduction band-edge in the different magnetic phases. The results of these calculations for bilayer \CPS\ are shown in Fig.~\figref{fig:5}{c}. The conduction band-edge in the ferromagnetic phase is indeed lower than in the antiferromagnetic one by approximately 30~meV. The result is robust because band structure calculations of multilayers from 2 to 6 layers and of bulk show that band edge shift is always there irrespective of thickness, and its order of magnitude remains the same (see a more detailed discussion in supporting information). These calculations also give the value of the effective mass --and hence of the density of states in the transistor channel-- from which the shift in band-edge energy $\Delta E$ can be estimated from the measured shift in accumulated charge density using the relation $\Delta n = \frac{m^*}{2\pi \hbar^2} \cdot \Delta E$ ($m^*=0.64 m_0$ is the effective mass of electrons in \CPS, $m_0$ is the free electron mass, and $\hbar$ is Planck's constant). We find $\Delta E \simeq 15$~meV using $\Delta n= C \cdot ( V_{\rm TH}(0~T) - V_{\rm TH}(8~T))$, a value that compares well to the calculated one, if the precision of the calculations to determine $\Delta E$ and the determination on the value estimated from the experiments are considered. This estimate is also consistent with the small threshold voltage shift observed for $T$ between 30~K and 50~K (Fig.~\figref{fig:3}{d}). That is because in that temperature interval, the smearing of the Fermi-Dirac distribution (close to 3.5~$k_B T$) ranges between 10 to 15~meV, and is comparable to the band edge shift. As a result of the limited resolution, the full shift in threshold voltage cannot be resolved.

The results of \emph{ab-initio} calculations can be qualitatively interpreted in a way that underscores the relation between the magnetic state of \CPS\ and its band structure. The conduction band-edge shift to lower energies originates from the increase of the bandwidth in the direction perpendicular to the layers. At zero applied magnetic field, adjacent layers have spin pointing in opposite directions and their energy differs by an amount close to the exchange energy, which is large (a fraction of 1~eV). Electrons are then forced to hop from one layer to the next-next layer, with an hopping amplitude that is exponentially suppressed due to the large distance involved. In the ferromagnetic state, instead, electrons can hop from one layer to the next, with the shorter distance that results in a substantially larger hopping energy, and therefore in a larger bandwidth in the \emph{c}-direction as compared to that in the antiferromagnetic state. The larger bandwidth implies that the states available for electrons start at lower energy, \ie\ that the conduction band-edge shifts to lower energy. Indeed, band structure calculations do show that the increase in bandwidth along the $c$-direction is responsible for the lowering of the conduction band-edge in the fully spin-aligned phase (see detailed discussion in supporting information S2). Clearly, as the magnetic field is increased, the transition from the situation corresponding to the antiferromagnetic state to that corresponding to the ferromagnetic one is gradual. Indeed --past the spin-flop transition field at 0.6~T-- the canting of the spins in \CPS\ occurs gradually up to approximately 8~T, at which point the spins are fully aligned at low temperature. The hopping integral in the $c$-direction --which is determined by the matrix element between spin states in adjacent layers-- therefore also increases gradually, and so does the bandwidth in the same direction. That is why the threshold voltage evolves gradually as the magnetic field is increased, before saturating above 8~T.

There are two important aspects of the experimental results presented here that should be retained. The first is that our results demonstrate that the band structure of \CPS\ depends on the magnetic state of the material. This is not obvious a priori, because --as we already mentioned in the introduction-- many 2D magnetic semiconductors investigated in the first generation of experiments (the Chromium trihalides, CrX$_3$ with X=Cl,~Br,~and~I~\cite{klein_probing_2018,song_giant_2018,wang_very_2018,kim_one_2018,ghazaryan_magnon-assisted_2018,klein_enhancement_2019,wang_determining_2019}, MnPS$_3$~\cite{long_persistence_2020}, VI$_3$~\cite{kong_vi3-new_2019,soler-delgado_probing_2022} and more) have extremely narrow bandwidths that cause their behavior to deviate from that expected for conventional semiconductors described by band theory. In \CPS\ the bandwidth is larger, approximately 1~eV (which is also why transistors work properly down to low temperature), and our results provide a first indication that band theory is suitable to describe the interplay between electronic and magnetic properties.

The second conclusion relates to the identified mechanism of magnetoconductance. A band shift had not been considered earlier as a possible cause of sizable magnetoconductance, likely because most magnetotransport studies on magnetic conductors are commonly performed on metallic systems, in which the effect is irrelevant because the Fermi level is located deep inside a band. For magnetic semiconductors the situation is different, because the Fermi level is commonly located near a band-edge, and that is why a band-edge shift can have such large effects. It seems clear that the mechanism identified here is not an exclusive property of \CPS\ but can play a role in many other magnetic semiconductors. For example, we expect that --under appropriate doping conditions-- the band shift associated to the reduction of the bandgap upon changing the magnetic state that has been reported in optical studies of CrSBr~\cite{wilson_interlayer_2021} (another 2D magnetic semiconductor) and of \ECA~\cite{santos-cottin_eucd_2as_2_2023} (a 3D magnetic semiconductor) may also manifest itself in the presence of a large magnetococonductance. The findings reported here therefore have much broader relevance than for the sole case of \CPS. 

\begin{acknowledgments}
The authors gratefully acknowledge Alexandre Ferreira for continuous and valuable technical support. We thank Dipankar Jana, Clément Faugeras, and Marek Potemski for fruitful discussion. AFM gratefully acknowledges the Swiss National Science Foundation and the EU Graphene Flagship project for support. MG acknowledges support from the Italian Ministry for University and Research through the Levi-Montalcini program and through the PNRR project ECS\_00000033\_ECOSISTER. K.W. and T.T. acknowledge support from JSPS KAKENHI (Grant Numbers 19H05790, 20H00354 and 21H05233).
\end{acknowledgments}

\FloatBarrier
%

\end{document}